# A Continent-Wide Assessment of Cyber Vulnerability Across Africa

Abdijabar Yussuf Mohamed

abdijabarmohamed3@gmail.com

Samuel Kang'ara Kamau
The Africa Digital Service

sam@africadigitalservice.org

## Abstract

As the internet penetration rate in Africa increases, so does the proliferation of the Internet of Things (IoT) devices. Along with this growth in internet access is the risk of cyberattacks to vulnerable IoT devices mushrooming in the African cyberspace. One way to determine IoT vulnerabilities is to find open ports within Africa's cyberspace. Our research leverages Shodan search engine, a powerful tool for discovering IoT devices facing the public internet, to find open ports across Africa. We conduct an analysis of our findings, ranking countries from most to least vulnerable to cyberattack. We find that South Africa, Tunisia, Morocco, Egypt, and Nigeria are the five countries most susceptible to cyberattack on the continent. Further, 69.8% of devices having one of the five most commonly open internet ports have had past documented vulnerabilities. Following our analysis, we conclude with policy recommendations for both the public and private sector.

## 1. Introduction

Despite technology remaining a bottleneck in its industrialization efforts, the African continent, home to approximately 1.4 billion people, is said to be on the cusp of colossal changes as the next frontier of the transformative digital boom. According to Internet World Stats estimations, as of December 31, 2021, Africa has 590.3 million internet users. This corresponds to an internet penetration rate of 43% and a growth rate of 12,975% from 2000 to 2021 (Internet World Stats, 2022). The corollary of an increased internet penetration rate is the rapid proliferation of the Internet of Things (IoT) into smart homes and critical infrastructures such as military applications, Industrial Control Systems (ICS), hospitals, and financial institutions. Nevertheless, the promise of a digital Africa has a sordid underbelly: every year, African economies lose millions of dollars to cybercrime.

As Africa digitizes, its critical infrastructure in cyberspace remains vulnerable to cyberattacks from rogue nation-states, international malicious actors, terrorists, and homegrown cybercrime syndicates. Arguably unbeknownst to many African stakeholders (governments, technology developers, and policy makers), there are many vulnerable devices across the African cyberspace. As per the 2021 African Cyberthreat Assessment Report by the International Criminal Police Organization (commonly known as Interpol), in 2021 alone, cybercrime reduced Africa's Gross Domestic Product (GDP) by more than 10%, which is equivalent to an estimated $4.12 billion (Interpol, 2021).

Despite the stakes involved, there is no research into the nature of vulnerable IoTs that face the public internet. Our research aims to conduct an in-depth study that clearly maps the vulnerable IoT devices in the entire region. Using Shodan, a search engine for indexing vulnerable IoT devices connected to the public internet, we unravel which ports in the Africa cyberspace are susceptible to being overtaken by malicious actors. Cybercriminals can leverage these vulnerabilities to gain unauthorized access into systems, plant malware into devices, and disrupt an organization's service by mounting Distributed Denial of Service (DDoS) attacks.

One of the ways criminals can gain unauthorized access into systems is through exploiting vulnerabilities in open ports that are listening for services. Before proceeding to the rest of our research, it is prudent we introduce the concept of ports and Shodan.

## 1. 1 Ports

A port is a virtual point where network connections start and end (Cloudflare, n.d.). These points are managed by a computer's operating system,which determines the kind of services each port operates. Operating a service includes either running a process or sending and receiving data packets from other ports. Ports enable computers to identify different kinds of network traffic. This then enables the routing of different services such as the delivery of a webpage's content and email delivery through dedicated ports. Figure 1 shows a diagram of a computer using port 443 to gain access to the internet through a HTTPS connection. The computer can also connect to the internet through port 80, meant for HTTP connections. Port 20 can be used for data transfers.

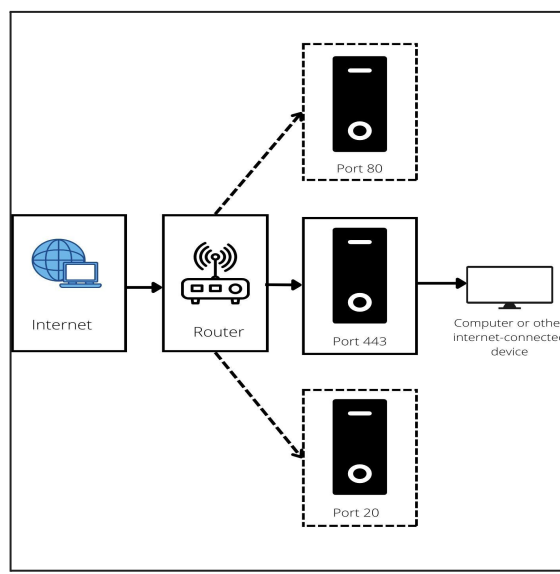
Fig. 1

Currently, most network applications use fixed port numbers assigned by the Internet Assigned Numbers Authority (IANA). The majority of internet applications included in these network applications use either of two key protocols to transmit their data packets: Transmission Control Protocol (TCP) or User Datagram Protocol (UDP). Open ports can receive data packets from other ports across the internet. Some ports are reserved for specific system functions and are therefore required to be open. Closed ports reject or ignore packets, but this also means they cannot receive data via packets from another port or device. The presence of either a TCP or a UDP port for these network applications implies the existence of a channel that can be exploited by a system intruder. Intruders can gain access to systems by exploiting common ports and sending attacks to the host through known system bugs. Intruders can also plant trojan horses through backdoors created via open port vulnerabilities. Through the use of Shodan, an Open Source Intelligence tool, we conducted a vulnerability assessment of open ports across Africa's cyberspace.

## 1.2 Shodan

In our daily browsing of the internet, we leverage web search engines such as Google Search and Microsoft Bing to search for content on the internet. Behind the scenes, search engines crawl the internet, index the content found during the crawling process, and render the relevant content to the user in a ranked order (Moz, n.d.). These search engines are used to find relevant content on websites. While the search engine Shodan shares parallels with these search engines, its fundamental difference is that it is a search engine for internet-connected devices.

Shodan can be described as a reconnaissance tool used by cybersecurity researchers and hackers, both ethical and malicious, to crawl the entire internet sphere and scan for IoT devices facing the public internet. It utilizes distributed port scanners throughout the globe to randomly select target Internet Protocol (IP) addresses and identify listening TCP and UDP ports. Due to the distributed nature of the port scanners, Shodan is able to search for open ports across all IP ranges and bypass any domestic or international bans for another country's IP ranges. Shodan randomly generates the IP addresses it scans rather than scanning through the addresses sequentially. This has been shown to ensure a uniform coverage of the internet and prevent bias in data collection (Matherly, 2015). The listening TCP and UDP ports are further enumerated to gather protocol banners, web pages, and other service data. In essence, Shodan's powerful algorithms grab the device/service banners and metadata that the server would normally send back to the client. On the one hand, banners refer to the textual information that describes the services running on a device. For web servers, banner information boils down to the headers that detail the type and version of the web server (Nginx, Apache, etc.) in place. On the other hand, metadata about the device entails such information as geographic location, Operating System (OS), hostname, and more (Matherly, 2015).

## 2. Data and Methodology

We used Shodan's search interface and Python to query results from the platform. Each Shodan search result includes the IP address of the host whose port is open. Each result also has metadata on the port including its function, name, location, and more. Shodan also makes its data available through API endpoints that serve individual results or bulk data. For each country in Africa, we gathered the distribution of open ports by their port numbers. We identified each open port's functions through its port number and metadata, thereby deciphering whether it is vulnerable to attack. Conducting this analysis on each country's open ports enabled us to compare each region's vulnerability to cyberattack. We analyzed data to find the most commonly open ports across Africa and each country's and region's top open ports. We also ranked the countries based on their degree of vulnerability to cyberattacks from most vulnerable to least vulnerable.

## 3. Results and Discussion

### 3.1 Most Commonly Open Ports

We find that connected devices across Africa are highly vulnerable to attacks. Three of the five most commonly open ports that devices across Africa use have documented vulnerabilities. We also find that 69.8% of the devices with one of the five most commonly open ports are vulnerable to cyberattacks.

As shown in Figure 2, the most open port is port number 7547, used by Internet Service Providers to communicate and update home routers. Infected routers account for 75% of attacks online (Norton, 2021).Vulnerabilities in this port have been used by malicious actors in the past to gain access to and control of the underlying device (Security Intelligence, 2016; C.P.S. Technologies, n.d.). Home routers are a particularly vulnerable attack vector because they function as a residential gateway, bridging a myriad of devices in households to the Wide Area Network or internet. This position means they can be used to access other devices connected to the router such as security cameras, webcams, smart TVs, and home devices such as Alexa. In avenues where these home routers with open ports are used to operate businesses, routers could be used to take control of the industrial control systems connected to them. In India, a vulnerability on port 7547 led to two state telecommunication companies having over 60,000 devices disconnected from the internet during an attack (Cimpanu, 2017). A similar attack in Germany in 2016 left 900,000 customers offline (Krebs on Security, 2016). Open ports on home routers can be used to collect user credentials, which then can be used to gain access to business systems or personal accounts. Credentials could include information such as credit card numbers, social media account login details, and even website or admin server access. These credentials could be used for identity theft and to access company information, such as customers' financial details, and even to access critical servers. Data on internet use, such as website and search histories, as recorded in the port traffic data would also leave users vulnerable to targeted or mass surveillance.

## 3.2 Top Open Ports per region

We analyze open port data across Africa's five regions: Northern Africa, Eastern Africa, Western Africa, Central Africa, and Southern Africa. We find that Northern and Southern Africa have the greatest number of connected devices with open ports that either have vulnerabilities or communicate through unencrypted channels that malicious actors can use to listen to messages being sent or received by the port.

Table 1 lists the most commonly open ports per region. The Eastern and Western Africa regions have fewer ports open as compared to each region's total population and total number of connected devices.

**Table 1**
*Top Open Ports Per Region*

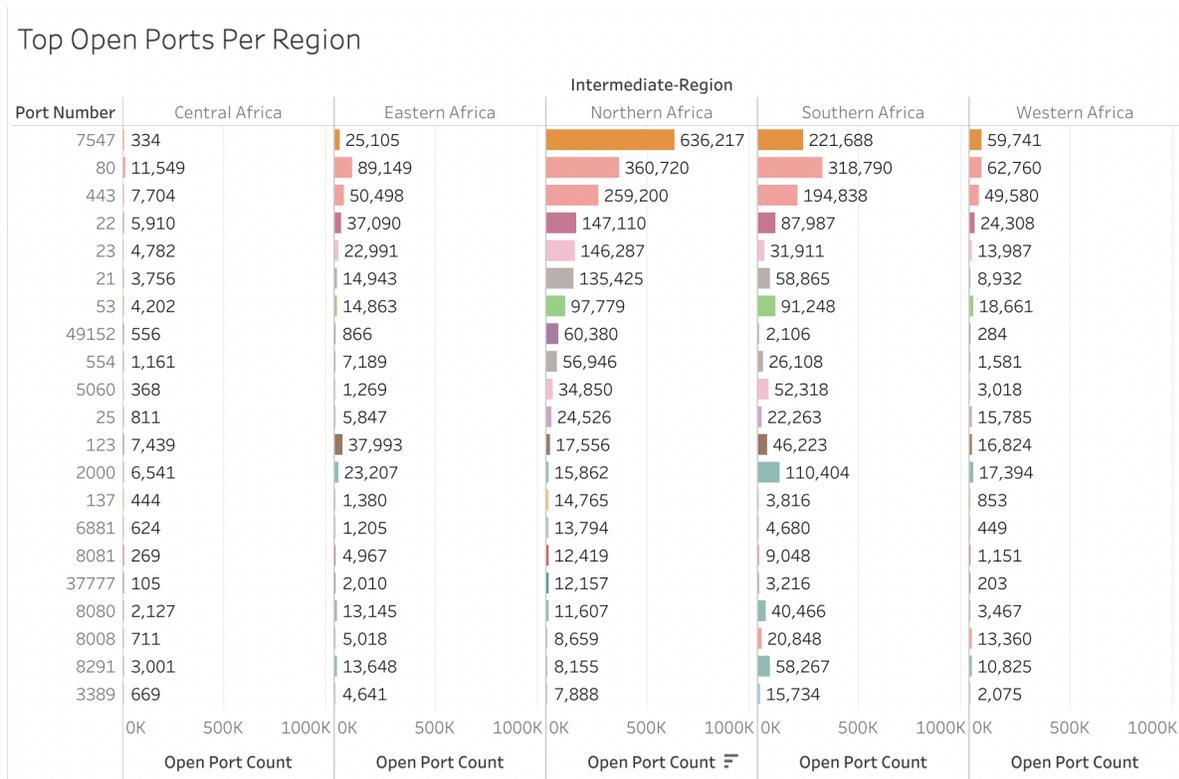

**Figure 2**

*Most Commonly Open Ports*

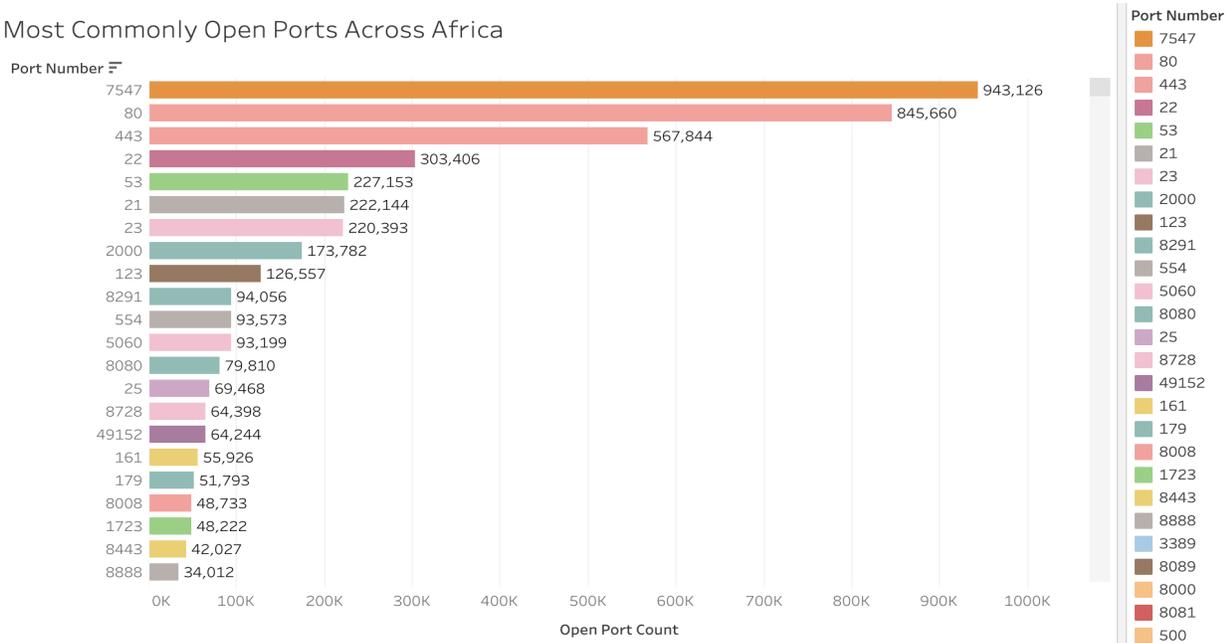

**Figure 3**

*Most Vulnerable African Countries by Open Internet Ports*

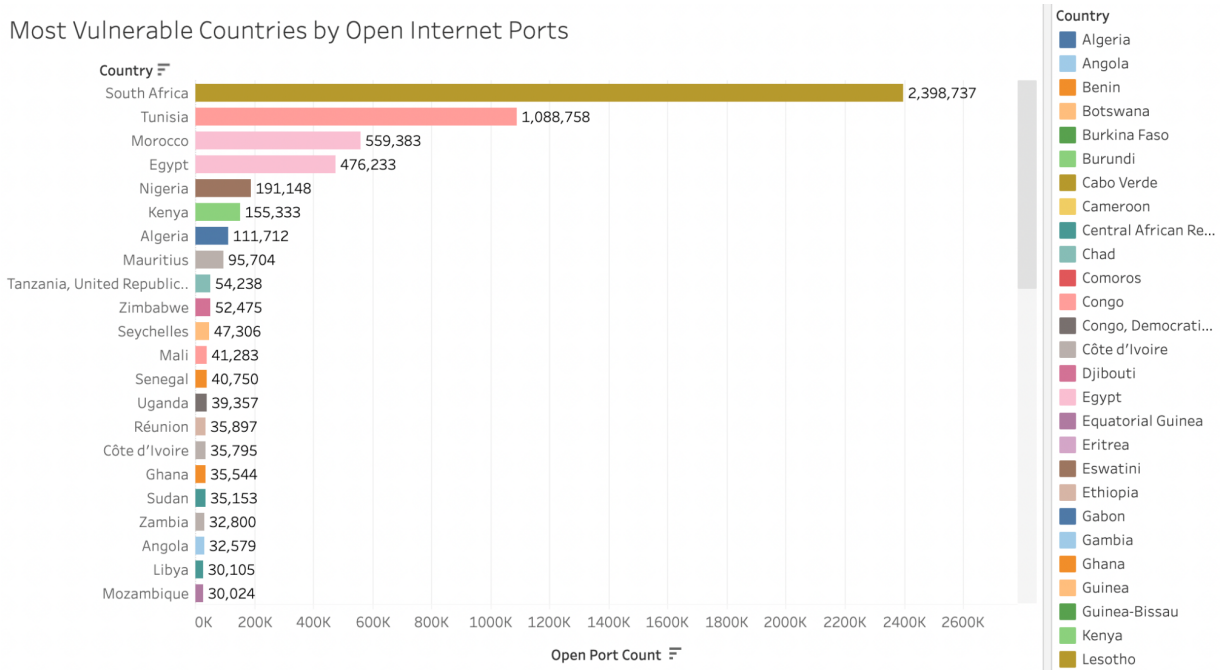

## 3.3 Countries with Most Open Ports

Matching the open port data to each country's metadata allowed us to find the number of open ports per country. This enabled us to analyze how vulnerable each country is to cyberattack. Depending on the networks a device is connected to, the unauthorized access of an open port in one country could lead to security vulnerabilities in other countries as well. For example, a cybercriminal gaining access to the device of a multinational institution might be able to access that organization's devices set up in another country either through backdoors, other forms of malware or social engineering. In Figure 3, we see that South Africa has the highest number of open ports. The previous regional breakdown suggests that these ports are mostly used for home routers, HTTP/HTTPS connections, remote network access, and FTP servers. Security vulnerabilities that can allow malicious actors to gain access and control of the host have been discovered in each of these ports apart from HTTPS (Security Intelligence, 2016; Abdulqader, 2016; McNulty, 2020).

## 3.4 Mapping Cyber Vulnerability Across Africa

Figure 4 shows a heatmap of cyber vulnerability across Africa. Some of the most vulnerable countries include South Africa, Egypt, Tunisia, Morocco, Nigeria, and Kenya.

Figure 4

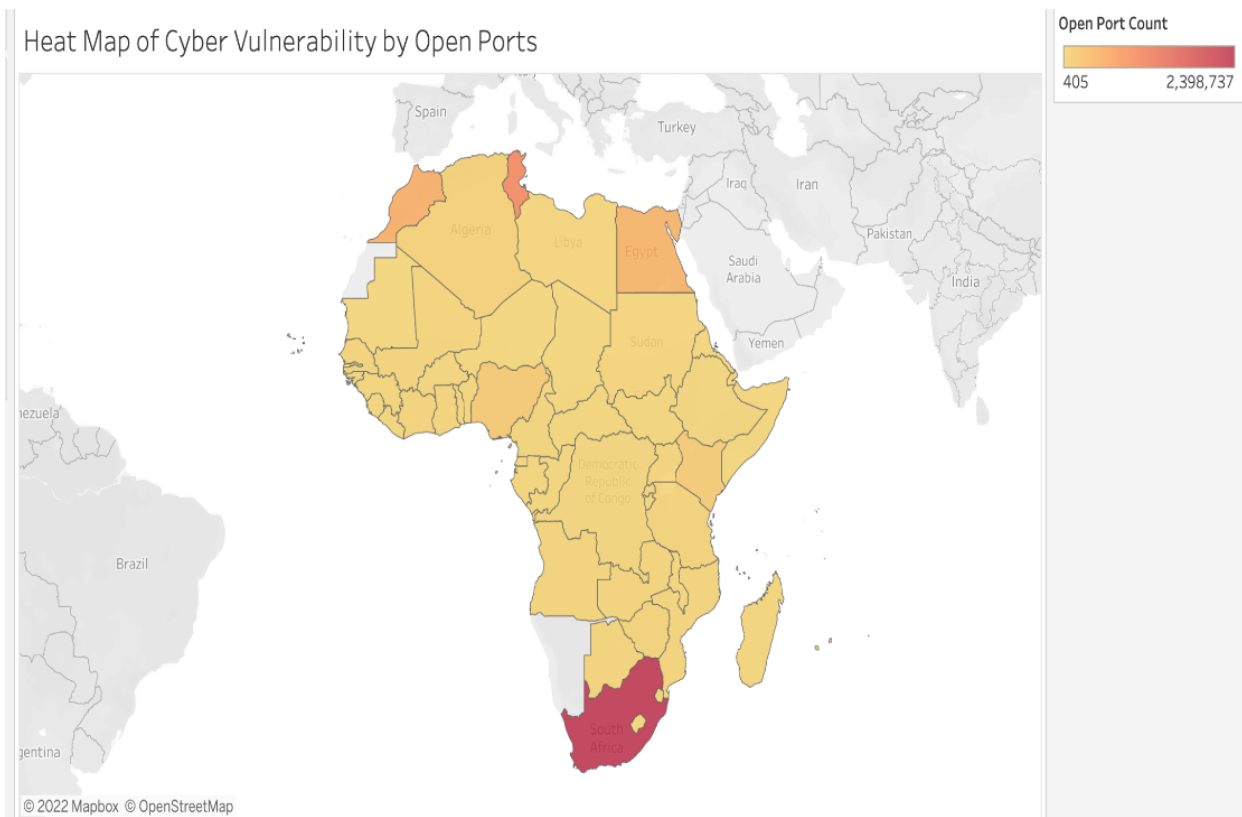

# 4.  Limitations and Further Work

Our study has taken steps to explore cyber vulnerability in Africa through synthesizing data on open ports across Africa's cyberspace. The current ranking of cyber vulnerability relies on a simple probabilistic model, where a higher number of open ports means more attack vectors for cybercriminals. A higher number of attack vectors means a greater likelihood of cybercriminals gaining unauthorized access to the open port's underlying systems. Thus, the more open ports a country has, the more likely it is to be vulnerable to cyberattacks. It would be interesting to also use data on historical attacks, such as the timeline and frequency of attacks on specific ports and devices. This would provide more detailed information on how susceptible each port is to attack according to historical data. Aggregating this data at the national level and comparing it to current open ports would enable researchers to better understand how two or more countries compare even when they have similar open port counts. Investigating country-specific cybersecurity policies would also enable researchers to identify countries with proactive or active responses to cyber threats as compared to those where little oversight on cybersecurity exists.

# 5.  Conclusion and Policy Recommendations

Using Open Source Intelligence tools, we identified open ports across Africa's connected devices. We further analyzed device metadata to conduct an assessment of each country's cyber vulnerability, ranking countries from most vulnerable to least vulnerable to cyberattacks. A complete list of rankings is available in Appendix A. We find that a significant number of the most commonly open ports on devices can be compromised. Because the most commonly open and vulnerable devices are home routers, the data raises privacy concerns for individuals and organizations alike. Cyberattacks are a real threat to African countries, and policy efforts are key to improving the resilience of connected devices. This in turn nurtures growing technology ecosystems and encourages the widespread adoption of technology, which is a key factor of economic growth.

In light of our research findings, we recommend that ISPs and organizations adopt policies that spearhead the implementation of the following key considerations:

### Patching vulnerable IoT devices

As soon as vulnerabilities are discovered on a given port or another service, it behooves the system administrators of organizations and the R&D teams at the ISPs to immediately conduct patch management, the process of distributing and applying updates to software. Patches are required in rectifying found vulnerabilities or bugs in a software. The quicker these patches are made, the more secure and less vulnerable the IoT devices would be.

### Port hardening

Internet-based communication occurs through ports; as such, for a given service to be rendered, such as sending and receiving an email, a specific port has to be open. However, when legitimate services are exploited through existing vulnerabilities, as can be done with Shodan, or malware is introduced to systems, malicious actors can leverage these services in conjunction with open ports to gain unauthorized access to critical and sensitive data. To prevent such a scenario, organizations and ISPs should close ports that are not associated with legitimate services, as open ports on a network increase the attack vectors available to malicious actors. This process, known as port hardening, includes other measures that ensure that the ports are not discoverable from the public internet.

### Guiding users to adopt best practice credential policies

Shodan contains default credentials (usernames and passwords) that can be used to gain access into systems. In combination with open ports, these can be used for nefarious purposes. Organizations are advised to guide their users to adopt best practice policy for selecting strong credentials. Furthermore, reinforcing that with multi-factor authentication could enhance the system's security.

### Regular network and system audits

Organizations could require their IT departments to conduct regular system audits. These can be used to find out whether the institution's systems have been compromised since their last check and could also identify unusual behavior on the system's traffic. These audits can run in real time and can also be automated to run at regular intervals and alert system administrators to unusual behavior on the network.

### Bounty programs

Institutions have seen success in establishing public bounty programs to help their engineering teams test software and identify system vulnerabilities. Bounty programs encourage cybersecurity professionals outside the organization to locate and file system vulnerabilities with an organization's engineering department for a reward. This would accelerate the identification of vulnerabilities before they are used to gain unlawful access into critical systems.

### Continental cybersecurity strategy

Previous recommendations have focused on policy practices at the organizational level. Currently, several countries in Africa have adopted national cybersecurity strategies (Government of Kenya, 2014; Government of South Africa, 2015). However, the fast-paced world of technology means that regulation is constantly catching up to the latest innovation. Adopting a continental cybersecurity policy framework under a common body such as the African Union would provide guidance to institutions operating in Africa on the baseline expectations for how they should set up their systems and handle user data. Similar policies such as the European Commission's General Data Protection Regulation have been implemented to protect users' online privacy ("General Data Protection Regulation (GDPR) – Official Legal Text," 2016).

## Acknowledgements

We would like to thank Stanford University's Digital Technologies in Emerging Countries (DTEC) initiative for supporting our research. Organized under the auspices of the Stanford Cyber Policy Center's Program on Democracy and the Internet (PDI) in partnership with the Center on Democracy, Development and the Rule of Law (CDDRL), the DTEC project seeks to explore the ramifications of emerging technologies on developing societies, democracies, and economies in the Global South, and provide evidence-based policy responses to these implications.

# Appendices

## A. Complete List of African Countries Ranked According to Cyber Vulnerability

| Ranking | Country | Open Port Count |
|---|---|---|
| 1 | South Africa | 2,398,737 |
| 2 | Tunisia | 1,088,758 |
| 3 | Morocco | 559,383 |
| 4 | Egypt | 476,233 |
| 5 | Nigeria | 191,148 |
| 6 | Kenya | 155,333 |
| 7 | Algeria | 111,712 |
| 8 | Mauritius | 95,704 |
| 9 | Tanzania, United Republic of | 54,238 |
| 10 | Zimbabwe | 52,475 |
| 11 | Seychelles | 47,306 |
| 12 | Mali | 41,283 |
| 13 | Senegal | 40,750 |
| 14 | Uganda | 39,357 |
| 15 | Réunion | 35,897 |
| 16 | Côte d'Ivoire | 35,795 |
| 17 | Ghana | 35,544 |
| 18 | Sudan | 35,153 |
| 19 | Zambia | 32,800 |
| 20 | Angola | 32,579 |
| 21 | Libya | 30,105 |
| 22 | Mozambique | 30,024 |
| 23 | Botswana | 26,645 |
| 24 | Cameroon | 22,637 |
| 25 | Eswatini | 19,374 |
| 26 | Burkina Faso | 15,293 |
| 27 | Cabo Verde | 14,855 |
| 28 | Ethiopia | 14,034 |
| 29 | Congo, Democratic Republic of the | 13,677 |
| 30 | Rwanda | 13,511 |
| 31 | Benin | 12,665 |
| 32 | Malawi | 11,852 |
| 33 | Madagascar | 11,843 |
| 34 | Togo | 10,943 |
| 35 | Lesotho | 7,535 |
| 36 | Gambia | 6,792 |
| 37 | South Sudan | 5,735 |
| 38 | Gabon | 5,320 |
| 39 | Guinea | 5,062 |
| 40 | Niger | 5,037 |
| 41 | Somalia | 4,977 |
| 42 | Burundi | 4,252 |
| 43 | Equatorial Guinea | 4,218 |
| 44 | Congo | 4,200 |
| 45 | Sierra Leone | 4,049 |
| 46 | Liberia | 3,963 |
| 47 | Chad | 3,554 |
| 48 | Mauritania | 3,136 |
| 49 | Djibouti | 1,585 |
| 50 | Sao Tome and Principe | 978 |
| 51 | Mayotte | 759 |
| 52 | Comoros | 688 |
| 53 | Central African Republic | 520 |
| 54 | Guinea-Bissau | 508 |
| 55 | Eritrea | 405 |

# B. List of Ports and their Functions

| Port # | Protocol | Description | Status |
|---|---|---|---|
| 0 | TCP, UDP | Reserved; do not use (but is a permissible source port value if the sending process does not expect messages in response) | Official |
| 1 | TCP, UDP | TCPMUX | Official |
| 5 | TCP, UDP | RJE (Remote Job Entry) | Official |
| 7 | TCP, UDP | ECHO protocol | Official |
| 9 | TCP, UDP | DISCARD protocol | Official |
| 11 | TCP, UDP | SYSTAT protocol | Official |
| 13 | TCP, UDP | DAYTIME protocol | Official |
| 17 | TCP, UDP | QOTD (Quote of the Day) protocol | Official |
| 18 | TCP, UDP | Message Send Protocol | Official |
| 19 | TCP, UDP | CHARGEN (Character Generator) protocol | Official |
| 20 | TCP | FTP Protocol (data) - port for transferring FTP data | Official |
| 21 | TCP | FTP Protocol (control) - port for FTP commands and flow control | Official |

| Port | Protocol | Description | Status |
|---|---|---|---|
| 22 | TCP, UDP | SSH (Secure Shell) - used for secure logins, file transfers (scp, sftp) and port forwarding | Official |
| 23 | TCP, UDP | Telnet protocol - unencrypted text communication, remote login service | Official |
| 25 | TCP, UDP | SMTP (Simple Mail Transport Protocol) - used for email routing between email servers | Official |
| 26 | TCP, UDP | RSFTP - A simple FTP-like protocol | Unofficial |
| 35 | TCP, UDP | QMS Magicolor 2 printer | Unofficial |
| 37 | TCP, UDP | TIME protocol | Official |
| 38 | TCP, UDP | Route Access Protocol | Official |
| 39 | TCP, UDP | Resource Location Protocol | Official |
| 41 | TCP, UDP | Graphics | Official |
| 42 | TCP, UDP | Host Name Server/WINS Replications | Official |
| 43 | TCP | WHOIS protocol | Official |
| 49 | TCP, UDP | TACACS Login Host protocol | Official |
| 53 | TCP, UDP | DNS (Domain Name | Official |

| | | System) | |
|---|---|---|---|
| 57 | TCP | MTP, Mail Transfer Protocol | Official |
| 67 | UDP | BOOTP (BootStrap Protocol) server; also used by DHCP | Official |
| 68 | UDP | BOOTP (BootStrap Protocol) client; also used by DHCP | Official |
| 69 | UDP | TFTP (Trivial File Transfer Protocol) | Official |
| 70 | TCP | Gopher protocol | Official |
| 79 | TCP | Finger protocol | Official |
| 80 | TCP | HTTP (HyperText Transfer Protocol) - used for transferring web pages | Official |
| 81 | TCP | Torpark - Onion routing ORport | Unofficial |
| 82 | UDP | Torpark - Control Port | Unofficial |
| 88 | TCP | Kerberos - authenticating agent | Official |
| 101 | TCP | HOSTNAME | |
| 102 | TCP | ISO-TSAP protocol/Microsoft | |

| | | Exchange | |
|---|---|---|---|
| 107 | TCP | Remote Telnet Service | |
| 109 | TCP | POP, Post Office Protocol, version 2 | |
| 110 | TCP | POP3 (Post Office Protocol version 3) - used for retrieving emails | Official |
| 111 | TCP, UDP | SUNRPC protocol | |
| 113 | TCP | Ident - old server identification system, still used by IRC servers to identify its users | Official |
| 115 | TCP | SFTP, Simple File Transfer Protocol | |
| 117 | TCP | UUCP-PATH | |
| 118 | TCP, UDP | SQL Services | Official |
| 119 | TCP | NNTP (Network News Transfer Protocol) - used for retrieving newsgroups messages | Official |
| 123 | UDP | NTP (Network Time Protocol) - used for time synchronization | Official |
| 135 | TCP, UDP | EPMAP / Microsoft RPC Locator Service | Official |
| 137 | TCP, UDP | NetBIOS NetBIOS Name Service | Official |

| Port | Protocol | Service | Status |
|---|---|---|---|
| 138 | TCP, UDP | NetBIOS NetBIOS Datagram Service | Official |
| 139 | TCP, UDP | NetBIOS NetBIOS Session Service | Official |
| 143 | TCP, UDP | IMAP4 (Internet Message Access Protocol 4) - used for retrieving emails | Official |
| 152 | TCP, UDP | BFTP, Background File Transfer Program | |
| 153 | TCP, UDP | SGMP, Simple Gateway Monitoring Protocol | |
| 156 | TCP, UDP | SQL Service | Official |
| 157 | TCP, UDP | KNET VM Command Message Protocol | |
| 158 | TCP, UDP | DMSP, Distributed Mail Service Protocol | |
| 159 | TCP, UDP | NSS-Routing | |
| 160 | TCP, UDP | SGMP-TRAPS | |
| 161 | TCP, UDP | SNMP (Simple Network Management Protocol) | Official |
| 162 | TCP, UDP | SNMPTRAP | Official |
| 170 | TCP | Print-srv | |
| 179 | TCP | BGP (Border Gateway Protocol) - an exterior gateway routing protocol that enables groups of | Official |
| | | routers to share routing information to ensure efficient and loop-free routes can be established. BGP is commonly used within and between ISPs. | |
| 190 | TCP, UDP | Gateway Access Control Protocol (GACP) | |
| 191 | TCP, UDP | Prospero Directory Service | |
| 192 | TCP, UDP | OSU Network Monitoring System, Apple AirPort Base Station PPP status or discovery, AirPort Admin Utility or Express Assistant | |
| 192 | TCP. UDP | SRMP (Spider Remote Monitoring Protocol) | |
| 194 | TCP | IRC (Internet Relay Chat) | Official |
| 201 | TCP, UDP | AppleTalk Routing Maintenance | |
| 209 | TCP, UDP | The Quick Mail Transfer Protocol | |
| 213 | TCP, UDP | IPX | Official |
| 218 | TCP, UDP | MPP, Message Posting Protocol | |
| 220 | TCP, UDP | IMAP, Interactive Mail AccessProtocol, version 3 | |
| 259 | TCP, UDP | ESRO, Efficient Short Remote Operations | |
| 264 | TCP, UDP | BGMP, Border Gateway Multicast Protocol | |
| 311 | TCP | Apple Server-Admin-Tool, Workgroup-Manager-Tool | |
| 318 | TCP, UDP | TSP, Time Stamp Protocol | |
| 323 | TCP, UDP | IMMP, Internet Message Mapping Protocol | |
| 383 | TCP, UDP | HP OpenView HTTPs Operations Agent | |
| 366 | TCP, UDP | SMTP, Simple Mail Transfer Protocol. On-Demand Mail Relay (ODMR) | |
| 369 | TCP, UDP | Rpc2portmap | Official |
| 371 | TCP, UDP | ClearCase albd | Official |
| 384 | TCP, UDP | A Remote Network Server System | |
| 387 | TCP, UDP | AURP, AppleTalk Update-Based Routing Protocol | |
| 389 | TCP, UDP | LDAP (Lightweight Directory Access Protocol) | Official |
| 401 | TCP, UDP | UPS Uninterruptible Power Supply | Official |
| 411 | TCP | Direct Connect Hub port | Unofficial |
| 427 | TCP, UDP | SLP (Service Location Protocol) | Official |

| Port | Protocol | Description | Status |
|---|---|---|---|
| 443 | TCP | HTTPS - HTTP Protocol over TLS/SSL (used for transferring web pages securely using encryption) | Official |
| 444 | TCP, UDP | SNPP, Simple Network Paging Protocol | |
| 445 | TCP | Microsoft-DS (Active Directory, Windows shares, Sasser worm, Agobot, Zobotworm) | Official |
| 445 | UDP | Microsoft-DS SMB file sharing | Official |
| 464 | TCP, UDP | Kerberos Change/Set password | Official |
| 465 | TCP | SMTP over SSL - CONFLICT with registered Cisco protocol | Conflict |
| 500 | TCP, UDP | ISAKMP, IKE-Internet Key Exchange | Official |
| 512 | TCP | exec, Remote Process Execution | |
| 512 | UDP | comsat, together with biff: notifies users of new c.q. yet unread e-mail | |
| 513 | TCP | Login | |
| 513 | UDP | Who | |
| 514 | TCP | rsh protocol - used to execute non-interactive commandline commands on a remote system and see the screen return | |
| 514 | UDP | syslog protocol - used for system logging | Official |
| 515 | TCP | Line Printer Daemon protocol - used in LPD printer servers | |

| Port | Protocol | Service | Status |
|---|---|---|---|
| 517 | TCP | Talk | |
| 518 | UDP | NTalk | |
| 520 | TCP | efs | |
| 520 | UDP | Routing - RIP | Official |
| 513 | UDP | Router | |
| 524 | TCP, UDP | NCP (NetWare Core Protocol) is used for a variety things such as access to primary NetWare server resources, Time Synchronization, etc. | Official |
| 525 | UDP | Timed, Timeserver | |
| 530 | TCP, UDP | RPC | Official |
| 531 | TCP, UDP | AOL Instant Messenger, IRC | |
| 532 | TCP | netnews | |
| 533 | UDP | netwall, For Emergency Broadcasts | |
| 540 | TCP | UUCP (Unix-to-Unix Copy Protocol) | |
| 542 | TCP, UDP | commerce (Commerce Applications) | |
| 543 | TCP | klogin, Kerberos login | |
| 544 | TCP | kshell, Kerberos Remote Shell | |
| 546 | TCP, UDP | DHCPv6 client | |
| 547 | TCP, UDP | DHCPv6 server | |
| 548 | TCP | AFP (Apple Filing Protocol) | |
| 550 | UDP | new-rwho, new-who | |
| 554 | TCP, UDP | RTSP (Real Time Streaming Protocol) | Official |
| 556 | TCP | Remotefs, rfs, rfs_server | |

| Port | Protocol | Description | Status |
|---|---|---|---|
| 560 | UDP | rmonitor, Remote Monitor | |
| 561 | UDP | monitor | |
| 561 | TCP, UDP | chcmd | |
| 563 | TCP, UDP | NNTP protocol over TLS/SSL (NNTPS) | Official |
| 587 | TCP | Email message submission (SMTP) (RFC 2476) | Official |
| 591 | TCP | FileMaker 6.0 Web Sharing (HTTP Alternate, see port 80) | Official |
| 593 | TCP, UDP | HTTP RPC Ep Map/Microsoft DCOM | Official |
| 604 | TCP | TUNNEL | |
| 631 | TCP, UDP | IPP, Internet Printing Protocol | |
| 636 | TCP, UDP | LDAP over SSL (encrypted transmission) | Official |
| 639 | TCP, UDP | MSDP, Multicast Source Discovery Protocol | |
| 646 | TCP | LDP, Label Distribution Protocol | |
| 647 | TCP | DHCP Failover Protocol | |
| 648 | TCP | RRP, Registry Registrar Protocol | |
| 652 | TCP | DTCP, Dynamic Tunnel Configuration Protocol | |
| 654 | TCP | AODV, Ad hoc On-Demand Distance Vector | |
| 665 | TCP | sun-dr, Remote Dynamic Reconfiguration | Unofficial |
| 666 | UDP | Doom, First online FPS | |
| 674 | TCP | ACAP, Application | |

| | | Configuration Access Protocol | |
|---|---|---|---|
| 691 | TCP | Microsoft Exchange Routing | Official |
| 692 | TCP | Hyperwave-ISP | |
| 695 | TCP | IEEE-MMS-SSL | |
| 698 | TCP | OLSR, Optimized Link State Routing | |
| 699 | TCP | Access Network | |
| 700 | TCP | EPP, Extensible Provisioning Protocol | |
| 701 | TCP | LMP, Link Management Protocol. | |
| 702 | TCP | IRIS over BEEP | |
| 706 | TCP | SILC, Secure Internet Live Conferencing | |
| 711 | TCP | TDP, Tag Distribution Protocol | |
| 712 | TCP | TBRPF, Topology Broadcast based on Reverse-Path Forwarding | |
| 720 | TCP | SMQP, Simple Message Queue Protocol | |
| 749 | TCP, UDP | kerberos-adm, Kerberos administration | |
| 750 | UDP | Kerberos version IV | |
| 782 | TCP | Conserver serial-console management server | |
| 829 | TCP | CMP (Certificate Management Protocol) | |
| 860 | TCP | iSCSI | |
| 873 | TCP | rsync - File | Official |

| | | synchronisation protocol | |
|---|---|---|---|
| 901 | TCP | Samba Web Administration Tool (SWAT) | Unofficial |
| 902 | | VMware Server | Unofficial |
| 911 | TCP | Network Console on Acid (NCA) - local tty redirection over OpenSSH | |
| 981 | TCP | SofaWare Technologies Remote HTTPS management for firewall devices running embedded Checkpoint Firewall-1 software | Unofficial |
| 989 | TCP, UDP | FTP Protocol (data) over TLS/SSL | Official |
| 990 | TCP, UDP | FTP Protocol (control) over TLS/SSL | Official |
| 991 | TCP, UDP | NAS (Netnews Admin System) | |
| 992 | TCP, UDP | Telnet protocol over TLS/SSL | Official |
| 993 | TCP | IMAP4 over SSL (encrypted transmission) | Official |
| 995 | TCP | POP3 over SSL (encrypted transmission) | Official |
| 3389 | (RDP) | Remote Desktop Protocol | |